# A state equation for the Schelling's segregation model

Jae Kyun Shin[a,*], Hiroki Sayama[b] and Seung Ryul Choi[a]

[a] School of Mechanical Engineering, Yeungnam University, Kyongsan 712-749, South Korea

[b] Collective Dynamics of Complex Systems Research Group, Binghamton University, State University of New York, Binghamton, NY 13902-6000, USA

[*]Correspondence author

## Abstract

An aspatial version for the famous Schelling's segregation model has recently been proposed, which, called two-room model, is still in an agent-based format like the original Schelling model. In the present study, we propose a new, state equation version of the Schelling model. The new equation is based on the two-room model and is derived in terms of a set of discrete maps. Fixed point solutions for the new equation are found analytically and confirmed numerically. Especially, we show that the extremely simple state equations can reasonably reveal the essence of the Schelling dynamics: integration, segregation and tipping. In addition to the fixed point solutions, periodic solutions are identified and conditions of the limit cycles are derived analytically.

Keywords : Schelling segregation, two-room model, population dynamics, state equation.

## 1. Introduction

We frequently find social or natural systems composed of two or more types of interacting groups of individuals. Cities composed of multiple races, ecological system composed of multiple species and social systems composed of individuals with different opinions, cultures, etc. Sometimes the interaction between different groups of agents gives rise to interesting global phenomena. Residential segregation in metropolitan cities, a sudden extinction of species in an ecological system and convergence to a unanimous opinion throughout the entire society are among the well known examples of the global phenomena.

To understand these phenomena, two different approaches have been in use: state equation based approach and agent-based one [1, 2]. The state equation based method is useful, in general, with relatively small number of equations making it easier to generalize the solutions. But with a limited number of parameters, it is not easy to reflect diverse differences between the individuals within the system. Agent-based models (ABM) could be a good alternative to address such issues. ABMs are highly flexible to incorporate detailed individual properties, spatially heterogeneous distributions and localized interactions of the agents in the model. However, the number of agents can be much larger than the number of corresponding state equations, making the solution process computationally expensive. In addition, we often need to simulate many times even for a given set of parameters, in order to draw a reliable conclusion from the numerical results. Even with such efforts, direct numerical simulation often provides limited analytical insight into the mechanisms underlying the observed phenomena. For this reason, the two methods are complementary. In many cases, both of these methods are well developed and selectively exploited depending on the specific target of the study.

The general comparison between the two types of models is valid also for the field of population dynamics of heterogeneous groups. Because of the flexibility of the ABMs, any state equation based model can be easily transformed into an ABM. We can find numerous ABMs for the Lotka-Volterra equations [3,4,5], for the epidemic spreading model [6,7] and for the opinion dynamics models [8-10]. But the inverse cases are rare. Finding a state equation for a given ABM will suggest a difficulty as the process usually requires some degree of abstractions. When a model is originally known only in terms of an ABM, it is challenging to derive its counterpart in state equation format. Schelling's model will be an example [11-14], for which only the agent-based format is known. Many other researchers modified the original Schelling model for a more realistic result, combining different factors affecting people's choice of their residences. However, most of these models are still ABMs and are fundamentally not different from the original Schelling model.

A couple of papers are found in the literature for treating the residential segregation problem using state equations [15,16]. Both of the papers are based on continuous differential equations in time and space. Especially, finding the density distribution of two types of agents on geometric space was the main target of these studies and diffusion is the key mechanism of change in these models. Although the method showed spatial segregation under some parameters, the results did not tell the key outcomes from the Schelling dynamics: segregation or integration as global phenomena. Diffusion type of equations has its limitation when applied for the Schelling problem, as the migration of the agents in the Schelling problem is clearly not continuous in space. Recently, we proposed the *two-room model* as an aspatial version of the Schelling problem[17]. Conceptually, the two-room model is still in agent-based format, and the solution process is computationally costly.

Motivated from the fact that there is no state equation version of the Schelling model, in the present study, we try to describe the Schelling problem in terms of a set of state equations. From the numerical results for the derived equations, we discuss the possibility and limitations of the state equations, when it is applied for the Schelling type problem. Also we discuss the oscillation mode obtained from the state equation.

## 2. Background: The two-room model

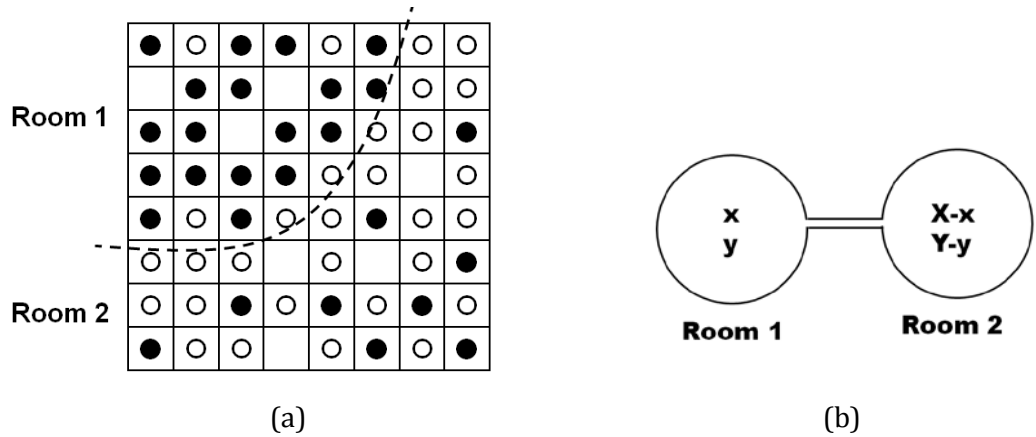

                     (a)                      (b)

**Fig. 1.** Two-room model[17].

    Consider the initial agent-based model of Shelling as shown in Fig. 1(a). Two types of agents, X an Y, are denoted as different types of circles. In the Schelling problem, different types represent, for example, White and Black residents. Initially the system is in a well mixed distribution so that every agent in the system has enough number of its same type neighbors. If there happen to be a disturbance that makes one of the agents unsatisfied, it will migrate to another place for a better neighborhood condition. A chain of migrations will follow until, possibly, a highly segregated region appear in the system as shown in Fig. 1(a), and eventually a complete segregation is obtained as the final outcome. Sudden change of the system from a well integrated state to a completely segregated one is known as *tipping*, especially when such a chain reaction is initiated by a slightest disturbance from the well integrated state.

    During the change in the system, such as toward the tipping, there appear different density zones for specific types of agents, as shown in Fig. 1(a). In such a case, we may be able to divide the system in two separate zones, or rooms, as shown in Fig. 1(a). Then the migrations of the agents will be explained as occurring between the two rooms as shown in Fig. 1(b). The agents can stay in one of the two rooms in the system, say room 1 and room 2, as shown in Fig. 1(b). Without ambiguity, the total population size of X agents will be denoted by X. The population size will be defined in a normalized sense. For example, X=1 and Y=0.5 means that the number or Y agents are half that of X. Same-group agents in the same room are assumed to be completely identical. Let us denote the number of

the type X agents in room 1 as x. In this case, the number of X agents in room 2 will be X-x. The only mechanism the system can change is through migration of agents between the two rooms. The total number and types of the agents does not change with the time. The question is how the agents are distributed in the two rooms. This will depend on the dynamics of the two types of agents.

## 3. State-equation version of the two-room model

In the original version of the two-room model was given in an agent-based format [17]. From now on, we will devise a new state-equation version of the original two-room model. The dynamics of the system can be described in terms of the rates of migration of the agents. The migration rate will be defined in terms of a difference in the potentials. We assume that the potentials depend only on the population contents of the rooms. If we denote the potential of an X-type agent using a function $\phi_x$, the potentials for each of the two rooms will be given as follows.

$$\text{Potential of an X in room 1} = \phi_x(x,y) \quad (3.a)$$
$$\text{Potential of an X in room 2} = \phi_x(X-x, Y-y) \quad (3.b)$$

No difference between the two rooms exist other than their population contents. Similarly the potential of the type Y agents can be described as $\phi_y(x,y)$, $\phi_y(X-x,Y-y)$ respectively. As in the ABM, the agents migrate in search of a room with a higher potential. Assuming the potential of the agents with Eq.(3) means that we are applying a concept similar to the mean field approximation, neglecting the spatial heterogeneity of the agents within a single room.

To determine the direction of migration, we use the difference in the potentials between the two rooms as in the followings.

$$\Delta\phi_x = \phi_x(x,y) - \phi_x(X-x, Y-y) \quad (4.a)$$
$$\Delta\phi_y = \phi_y(x,y) - \phi_y(X-x, Y-y) \quad (4.b)$$

If $\Delta\phi_x$ is positive, it means that room 1 is more attractive than room 2 for X. In such a case, the agents in room 2 will try to migrate to the room 1. The rate of migration can be reasonably assumed to be proportional to the potential difference and the number of agents in room 2. As a result, the rate of migration, $r_n$, of X agents per unit time step will be described as follows.

$$r_n = \begin{cases} k_x \Delta\phi_x (X - x_n) & \text{when } \Delta\phi_x \geq 0 \quad (5.a) \\ k_x \Delta\phi_x x_n & \text{when } \Delta\phi_x < 0 \quad (5.b) \end{cases}$$

Equation (5.a) defines the migration rate $r_n$ from room 2 to room 1, as proportional both to the potential difference, $\Delta\phi_x$, and to the number of the agents in room 2, $X-x_n$. If the potential difference is negative, the migration will take place in the opposite direction as described in Eq. (5.b). The subscripts *n* denotes the current time step and $k_x$ denotes a constant.

From the above equations, we can get a discrete map version of the two-room model as follows.

$$x_{n+1} = \begin{cases} x_n + k_x \Delta\phi_x (X - x_n) & \text{when } \Delta\phi_x \geq 0 \\ x_n + k_x \Delta\phi_x x_n & \text{when } \Delta\phi_x < 0 \end{cases} \quad (6.a)$$

$$y_{n+1} = \begin{cases} y_n + k_y \Delta\phi_y (Y - y_n) & \text{when } \Delta\phi_y \geq 0 \\ y_n + k_y \Delta\phi_y y_n & \text{when } \Delta\phi_y < 0 \end{cases} \quad (6.b)$$

## 4. Linear potential functions

The dynamics of the two room model will depend on the potential functions $\phi_x(x,y)$ and $\phi_y(x,y)$. For the present study, we will assume that both of the preference functions are linear, as in Eq.(7) below.

$\phi_x (x,y) = ax+by$     (7.a)

$\phi_y (x,y) = cx+dy$     (7.b)

Here, a and d denote intra-group preference constants and b and c, inter-group constants. In Schelling-type problems, each of the groups is assumed to prefer same-type neighbors. Such a situation can be described by choosing a>b and d>c. The linear relationship shown in Eq.(7) may not be good when we consider the original Schelling model, where an agent is satisfied when the fraction of the same-type neighbors is above some threshold [12,17]. The constants may have to be step functions in order to reflect the threshold neighborhood demand more precisely. However, linear approximation makes the equation easier to analyze and may be good as a first approximation. In fact, linear approximation was sometimes used for the Schelling problem [18,19]. We define potential differences f and g, for the linear case, as follows:

$$f(x,y) \equiv \Delta\phi_x = 2ax + 2by - aX - bY \qquad (8.a)$$

$$g(x,y) \equiv \Delta\phi_y = 2cx + 2dy - cX - dY \qquad (8.b)$$

Substituting Eq.(8) into Eq.(6), we obtain the iterative maps for the linear case as follows:

$$x_{n+1} = \begin{cases} x_n + k_x(2ax_n + 2by_n - aX - bY)(X - x_n) & \text{when } f(x_n, y_n) \geq 0 \\ x_n + k_x(2ax_n + 2by_n - aX - bY)x_n & \text{when } f(x_n, y_n) < 0 \end{cases} \qquad (9.a)$$

$$y_{n+1} = \begin{cases} y_n + k_y(2cx_n + 2dy_n - cX - dY)(Y - y_n) & \text{when } g(x_n, y_n) \geq 0 \\ y_n + k_y(2cx_n + 2dy_n - cX - dY)y_n & \text{when } g(x_n, y_n) < 0 \end{cases} \qquad (9.b)$$

The system in Eq. (9) is a set of quadratic maps with discontinuity in the slope of the rate functions at $f = 0$ and $g = 0$. General analytic solutions may not be available for the system. But we have no difficulty in solving the system numerically. If we maintain $k_x$ and $k_y$ sufficiently small, the system remains in $0 \leq x \leq X, 0 \leq y \leq Y$, for any initial values of $0 \leq x_0 \leq X, 0 \leq y_0 \leq Y$. For simplicity, we used $k_x = k_y = 0.1$ throughout the present study.

We will investigate the dynamics of the system for two purposes in this paper. In this section, we will see possible types of solutions for the case with X=Y=1. By setting $x_{n+1}=x_n$ and $y_{n+1}=y_n$ simultaneously in Eq.(8), we can easily find a set of fixed points as listed in Table 1. There can be total of nine fixed points in the system. In many cases, the fixed solutions NO. 5 to NO. 9 shown in the Table 1 are unstable solutions. These will be explained later for example cases. Because the two rooms are not distinguishable, the points (0, 0) and (1,1) means essentially the same thing, the integration(INT) of the two types of agents in one room. Similarly both (0,1) and (1,0) mean segregation(SEG). The type of solutions actually obtained from the numerical simulation depends on the parameter set and the initial conditions. The point (1/2, 1/2) is of special interest. This state means a completely integrated state (denoted by INT2) for which the populations are evenly distributed between the two rooms. However, in most cases, the state is unstable and a slight disturbance from INT2 will drive the system into segregation or another type of integration, (0,0) or (1,1). The INT2 mode is found to be not a stable outcome of the system, except for some carefully designed parameter sets. So we neglect the importance of the mode as a fixed solution. INT2 can be useful only as an initial condition to check the behavior of the system. Using slightly disturbed initial conditions from the unstable mixed state of INT2, we can observe the outcome of the system, INT or SEG, for example.

Time series solutions converging two of the fixed points, representing segregation and integration, respectively, are shown in Fig. 2(a) and (b), as examples. In addition to these fixed points, limit cycles can be obtained, an example of which is shown in Fig. 2(c).

Table 1 Possible set of fixed points

| NO | (x,y) | (When X=Y=1) | Type(Figs.3, 4) |
|---|---|---|---|
| 1,2 | (0,0),(X,Y) | (0,0),(1,1) | INT |
| 3,4 | (0,Y),(X,0) | (0,1),(1,0) | SEG |
| 5 | (x*,y*) | (1/2,1/2) | INT2 |
| 6 | (0.5X+0.5bY/a,0) | (0.5+0.5b/a,0) | UNSTABLE |
| 7 | (0.5X-0.5bY/a,Y) | (0.5-0.5b/a,1) | UNSTABLE |
| 8 | (0,0.5Y+0.5cX/d) | (0,0.5+0.5c/d) | UNSTABLE |
| 9 | (X,0.5Y-0.5cX/d) | (1,0.5-0.5c/d) | UNSTABLE |

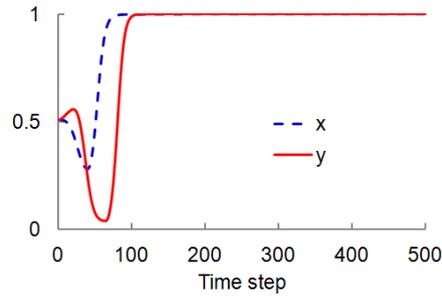

(a) (b,c)=(-0.9,1.1)

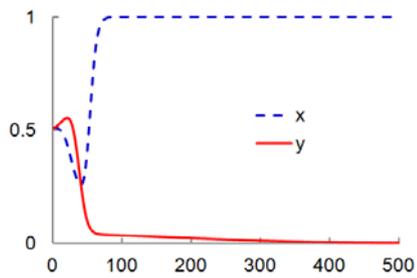

(b) (b,c)=(-1.0,0.9)

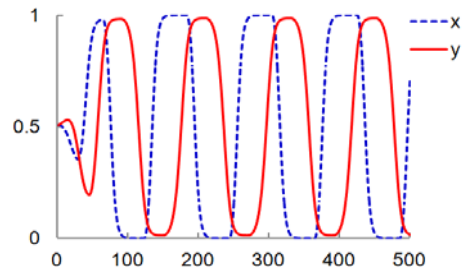

(c) (b,c)=(-1.5,1.0)

**Fig. 2.** Types of solutions with initial values $(x_0,y_0)=(0.51,0.51), a=d=1.0$: (a) Integration, (b) Segregation, (c) Oscillation (Limit cycle).

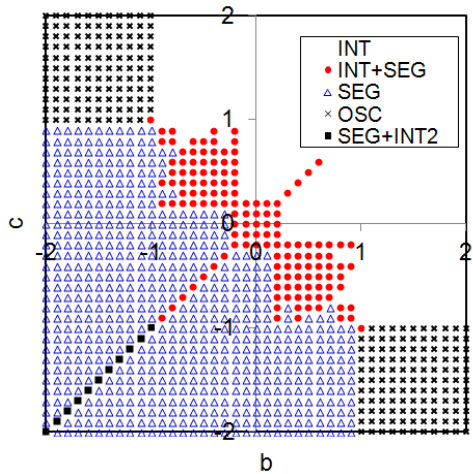

**Fig. 3.** Phase diagram. For $a=d=1.0$ and $X=Y=1.0$. Modes classified using initial condition sets (0.51,0.51), (0.51,0.49), (0.49,0.51) and (0.49,0.49).

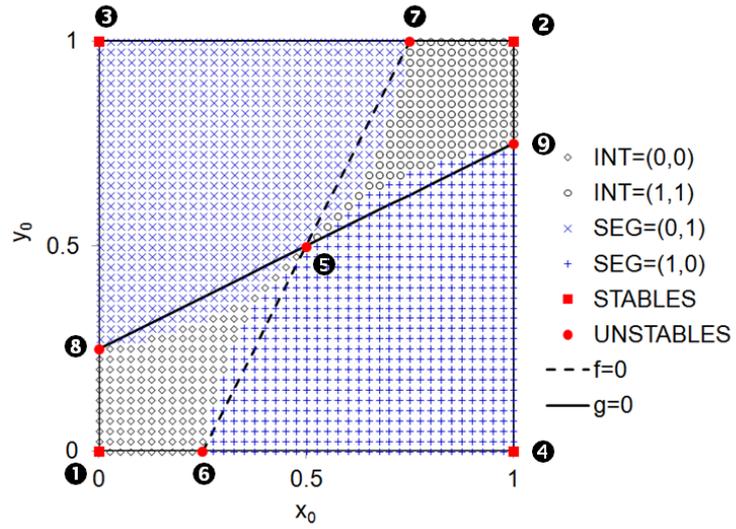

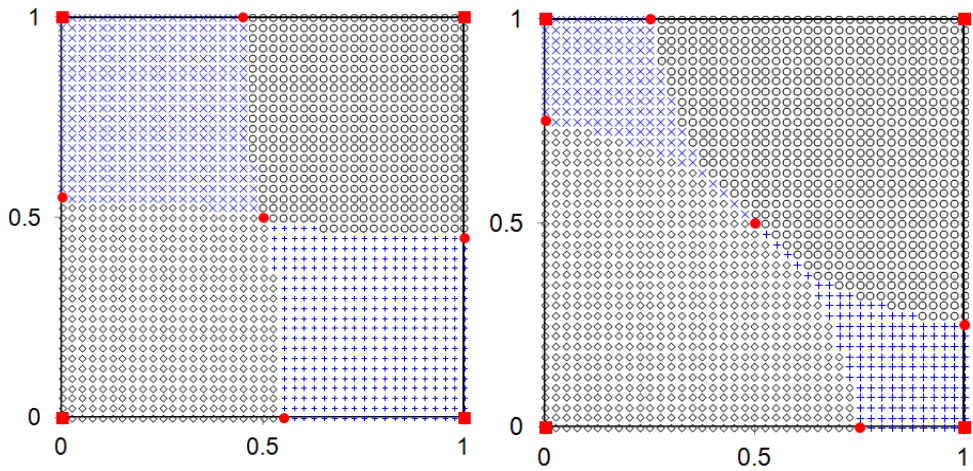

**Fig. 4.** Basins of attraction for (a)b=c=-0.5, (b)b=c=0.1, (c)b=c=0.5. Circled numbers in (a) corresponds to the solution number shown in Table 1.

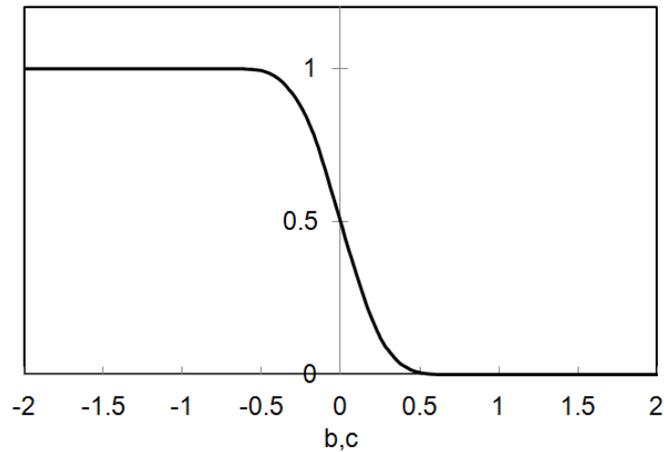

**Fig. 5.** Segregation probability( $a=d=1$ ). Initial conditions $(x_0, y_0)=(0.5+\rho \cos \theta, 0.5+\rho \sin \theta)$, both $\rho$ and $\theta$ uniform random, $0 \leq \rho \leq 0.1$, $0 \leq \theta \leq 2\pi$.

      A phase diagram Fig. 3 shows the type of solutions obtained from the system. For Fig.3, we fixed the parameters $a=d=1.0$ and the phase diagram is plotted on (b, c) plane on the intervals $-2 \leq b \leq 2$, $-2 \leq c \leq 2$. For the phase diagram, solution type at each of the points is classified from the four solutions obtained with four different initial conditions, (0.51, 0.51), (0.51, 0.49), (0.49, 0.51) and (0.49, 0.49). These initial conditions are selected to simulate the system which is slightly perturbed from a perfectly mixed state of $(x, y)=(0.5, 0.5)$. The basins of attraction sampled in Fig. 4 show that the said four initial conditions will be good for checking the possible type of solutions for a given parameter set b and c.

      Types of the solutions in Fig. 3 are classified into five groups. Most of the first quadrant of the parametric space is occupied by the integration (INT) mode. In this zone, only the fixed points (1, 1) or (0, 0) are obtained without regard to the initial conditions. For the first quadrant, when b and c is positive and not too small in magnitude, meaning that both X and Y being quite attractive to each other, integration will be a natural consequence. Likewise, most of the third quadrant is occupied by segregation(SEG) mode. With b<0 and c<0 at the third quadrant, the two types repel each other and segregation will be the final outcome. Far sides of the second and fourth quadrants are occupied by oscillation (OSC) mode. The solution is given not as fixed points but as limit cycles, as exampled in Fig. 2(c). In these zones, the two types of agents behave like chaser and runner. For example, in the second quadrant, we have b<0 and c>0. In this case, X repels Y but Y will chase X. X will be a runner and Y, a chaser. On the fourth quadrant, the roles are reversed. The oscillation mode is an interesting type of motion that was not possible in the original Schelling model. In the

next section, the oscillation mode will be analyzed in detail. In the intermediate zone between the three modes lies a mixed mode (INT+SEG) of integration and segregation. The system can go either integration or segregation, depending on the initial conditions. All the parameter sets sampled for the Fig. 4 belongs to the mixed mode. It is interesting that the segregation mode is possible even for the case b>0 and c>0.

For all the case in the Fig. 4, marked are five unstable fixed points, respectively. The possible existence of these five unstable fixed points were already predicted in Table 1 and indeed found numerically. In Fig 4(a), the locations of all the nine fixed point solutions are marked with numbers. Each of the unstable fixed points can be obtained when the initial condition $(x_0, y_0)$ is exactly the same as the fixed solution itself.

The mixed mode is interesting when we consider the Schelling dynamics. We can conclude that the tipping in the Schelling dynamics is possible only for the mixed zone and the SEG zone. Only in these two zones, the system can go from a perfectly mixed state of (0.5, 0.5) to a segregated state, when a slightest disturbance is applied. What is really interesting is such a tipping is possible even in the first quadrant, where b>0 and c>0. This means that the segregation is possible even if X and Y are attractive to each other, when the inter-group attractiveness (b, c) is relatively smaller than the intra-group one (a,d).

To observe the parameter range for the possible tipping, segregation probability is calculated. To simplify the discussion, only symmetric parameter sets, a=d=1 and b=c, are used. In most of the ABM type Schelling models, as well as in the original Schelling model, symmetric parameter sets are used. These symmetric parameter sets lie on the diagonal line crossing the first and third quadrant in Fig. 3. For a given set of parameters, we conducted a Monte Carlo simulation with initial conditions randomly lying on the circle of radius of 0.1 and center at (0.5, 0.5). From the simulation we can obtain the probability the system can show SEG with a random slight disturbance from (0.5, 0.5). The probability of segregation thus obtained is plotted in Fig. 5. Below b=c=-0.5, segregation is the only possible mode. The probability decreases from b=c=-0.5 until it becomes essentially zero at b=c=0.5, beyond which point only the integration mode is possible. Between -0.5 to 0.5, both integration and segregation is possible depending on the initial conditions. When both integration and segregation is possible, we should conclude that, in the event, the system will converge toward the segregation. In the two-room model, we have two types of integration, (0, 0) type or (1/2,1/2) type. When we consider the Schelling dynamics, these two types of integration, INT and INT2, will mean the same, as there can be only one type of integration in reality.

Thus, when we started from an initial condition near the point (0.5, 0.5) and get fixed point solutions like (0,0) or (1,1), this should be interpreted that the system has not changed from the initial state. When we get (0,1) or (1,0) as a final outcome, the system has gone tipping from an integration toward a segregation. Thus, when b=c<0.5, the segregation will be the only outcome of the system in the long run. In conclusion, we can say that the system can tip when the inter-group preferences (b,c) falls below half the intra-group preferences (a,d). From this, we can define the critical intergroup preferences as $b_{crit}=c_{crit}=0.5$. The critical value can be compared to the critical neighborhood demand in the original Schelling model[12]. The two-room model with linear potential functions can reveal the essence of the Schelling dynamics, at least in a qualitative sense. Quantitatively, however, the results from the two-room model cannot be directly compared to the ABM. Future study is necessary, for example, to convert the neighborhood demand into the constants a~d more realistically.

In the phase diagram of Fig.3, there is fifth mode marked SEG+INT2. But Fig. 5 shows that the probability to get an INT2 solution will be almost zero because the segregation probability is essentially 1.0 at b=c<-1.0. As explained before, the probability to obtain INT2 mode is almost zero as it can be obtained only at initial conditions $x_0=y_0$.

## 5. Oscillation mode

From the phase diagram of Fig.3, we saw that the oscillations can be obtained in the second and the fourth quadrants. In this section, we will analyze the oscillation mode in detail. First, the conditions of the oscillation mode will be derived and next parametric dependence of the oscillation frequency will be studied. Derivation of the condition of limit cycle starts from the fact there can be two types of fixed points, integration and segregation. If the system is to be in an oscillation mode, it is necessary that the system should not be attractive to any of these fixed points. Let us first consider a condition that the system should not stay fixed at the integrated state of, for example, (x,y)=(X,Y). In such an integrated situation, the room 2 will be empty and the potential of room 2 for both types of agents will be 0. If oscillation is to be obtained, the potential is such that the type X agents should prefer room 2, meaning that the potential of room 1 should be less than 0 for the X type agents. The condition can be written as, aX+bY<0. Or

$$b < -a\frac{X}{Y} \qquad (10)$$

Similarly, we can find the condition that the fixed state of (X,0) or (0,Y) cannot be maintained, as is given in Eq.(11) below.

$$c > d\frac{Y}{X} \qquad (11)$$

Although inequalities (10) and (11) are not necessary conditions, it works quite well at least in the case shown in Fig. 3. When we use X=Y=1 and a =d=1 as in Fig. 3, we get the said conditions as b<-1 and c>1, which seems to be good as can be observed in Fig. 3.

As stated in the previous section, the oscillation mode occurs when the two types of agent behaves like chaser and runner. In the real world, dynamics between chaser and runner type of agents could be found in the predator-prey model [4]. Each of the predators and preys tends to gather together, meaning a, d>0. For the predators, the preys will be attractive (b>0) but the preys will escape from the predators(c<0). The parameters can satisfy the oscillation condition. However, in the predator-prey models in the literature, the main interest lies on the change in the size of the populations, such as distinction or persistence of the species. In those models, predation and regeneration is more important than the migration, the latter being the only mechanism in the present two-room model. So the oscillation is not realized in real predator-prey model.

We found an interesting phenomenon in the literature that shows an oscillation mode: rapid pole-to-pole (PP) oscillation of the bacteria E. coli [20-21]. During a cell division process of E. Coli, two types of proteins, minD and minE, rapidly oscillate between the two poles of the cell. The phase lag found in the present oscillation mode is similarly observed between the migrations of the minD and minE, minD leading the phase. In the Schelling model, the rooms were conceptually divided as explained in Fig. 1 and does not necessarily exist in physical domain. But in the PP oscillation, the two poles take the role of the two rooms and the rooms really exist in the physical domain. In the following, we will investigate the oscillation mode in detail, assuming that we are trying to explain the PP oscillation in terms of the two-room model. We are not telling that the minD and minE really behave like runner and chaser, respectively. Nor we tell that the two-room model can be really a valid model for the PP oscillation. We just use the PP oscillation as a guideline for our parametric study.

In the bacterial cell division process discussed, it is reported that the oscillations occur only for a certain density range of minE proteins for a fixed density of minD proteins [20,21]. In other words, there is a threshold density of MinE above which the oscillation is possible and a limit beyond which the oscillation disappears. In terms of the present paper, it is the same as saying that the oscillation can happen only for a certain range of Y if X is kept constant. Figure 6 explains that this will be really the case here. The parameter zones for the oscillation mode are drawn for three different values of the total population Y, while keeping X=1. The boxes are cut at b=-3 and c=3. These boxes can be easily obtained from the necessary conditions, inequalities (10) and (11). To see the dependence of the oscillation on the relative population size, take the parameter set, as an example, at b=-2.0 and c=2.0, which is marked by a circular dot in Fig. 6. The mark is clearly inside of the oscillation zone for X=Y=1. However, the same parameter set lies on the border line for each of the oscillation zones for Y=2.0 and 0.5. From Fig. 6, we can easily predict that the point will lie outside of the oscillation zone for Y>2.0 or Y<0.5. For the assumed set of parameters, we can conclude that the oscillation mode will be possible only for 0.5<Y<2.0 for X=1.0. This kind of logic will be valid, in general, for other parameter sets, although the limiting values will depend on the specific parameter set.

The actual range of Y values for the oscillation mode can be directly obtained through numerical simulations of the two-room model. The results are shown in Fig. 7. In Fig. 7, the frequency of oscillation is plotted as a function of Y for selected values of X. Oscillation is obtained only for the range of Y values the graph is plotted. We can observe that the range is between Y=0.5 and 2.0 for the case X=1, which is in agreement with the range predicted indirectly using the necessary conditions. The graph shows that the period of oscillation, in general, decreases with the increase in Y. The dependence of the oscillation frequency on the relative population size again shows similar behavior with the PP oscillation [20]. However, we again stress that these similar behavior does not mean that the two-room model is really valid for explaining the PP oscillation. Precise quantitative comparison of the current result with the experimental one from the PP oscillation will be a topic of future study and is beyond the scope of the current work.

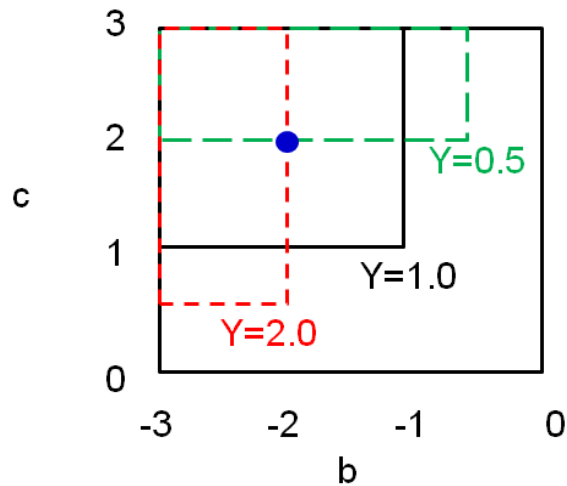

**Fig. 6.** Oscillation zones for different values of Y.(X=1)

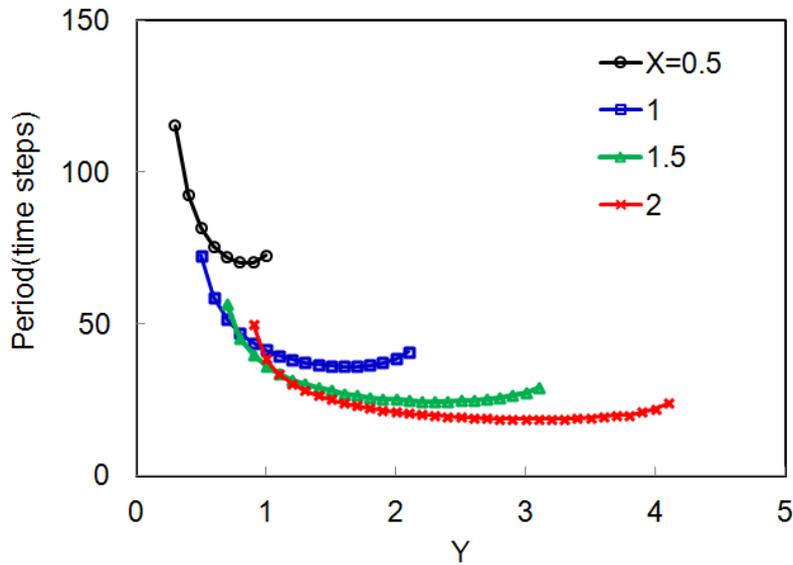

**Fig. 7.** Period of oscillation depending on population size. (b,c)=(-2.0,2.0).

## 6. Conclusions

We proposed a state equation version of the two-room model which was initially known as an agent-based, aspatial version of the Schelling's segregation model. The state equation is given as a set of discrete maps with a set of four parameters. The parameters are meant to reflect the neighborhood preferences of the original Schelling model [12]. It may be needless to say that such a simple set of equations cannot fully replace the original spatial, agent-based model. Between the two models, there is a fundamental difference in defining the level of satisfaction of an agent. In the original Schelling model, the agents' estimation is based on local information. Only the agent's immediate neighbors contribute to the focal agent's satisfaction. But in the present model, all the population in the system contributes to it. In addition, a mean-filed concept is incorporated in the present model and, as a result, the spatial heterogeneity, which is very important in the original agent based model, is neglected. The spatial heterogeneity is minimally considered by introducing two separate rooms not identical in their densities in specific type of agents. In spite of its simplicity with only a set of four parameters, however, the proposed model revealed key dynamic behavior of the Schelling dynamics, the integration and segregation. Especially, the possibility of the tipping, which will be the symbol of the Schelling dynamics, was predicted from the existence of the INT+SEG and SEG zone in the parametric space. In addition to the integration and segregation mode, an oscillation mode is indentified when the parameter sets indicate a runner-chaser relation for the two populations. Conditions for the oscillation mode was suggested and numerically identified.